\documentclass[aps,prd,amsmath,amssymb,superscriptaddress,altaffillsymbol, preprintnumbers,preprint,nofootinbib,a4paper]{revtex4}

\pdfoutput=1
\usepackage{amsthm}
\usepackage{graphicx}
\usepackage{color}
\newcommand{\beq}{\begin{eqnarray}}
\newcommand{\eeq}{\end{eqnarray}}

\newcommand{\bmp}{\noindent\begin{minipage}{16cm}}
\newcommand{\emp}{\end{minipage}\vskip 7mm}

\usepackage{dcolumn}
\usepackage{bm}
\usepackage{bbm}
\usepackage{subfigure}
\usepackage{pxfonts}
\usepackage{slashed}

\theoremstyle{definition}

\theoremstyle{plain}

\usepackage{epsfig}

\usepackage[ margin=5pt, font=normalsize,labelfont=bf,justification=raggedright]{caption}

\usepackage{hyperref}
\definecolor{rossoCP3}{cmyk}{0,.88,.77,.40}
\definecolor{verdeCP3}{rgb}{0.09765625, 0.57421875, 0.1015625}
\definecolor{bluCP3}{rgb}{0, 0.23, 0.67}
\hypersetup{colorlinks, bookmarksopen, bookmarksnumbered,citecolor=verdeCP3, linkcolor=bluCP3, pdfstartview=FitH, urlcolor=rossoCP3}

\def\lsim{\mathrel{\rlap{\lower4pt\hbox{\hskip1pt$\sim$}}
    \raise1pt\hbox{$<$}}}                
\def\gsim{\mathrel{\rlap{\lower4pt\hbox{\hskip1pt$\sim$}}
    \raise1pt\hbox{$>$}}}

\newcommand{\bea}{\begin{eqnarray}}
\newcommand{\eea}{\end{eqnarray}}

\newcommand{\ba}{\begin{eqnarray}}
\newcommand{\ea}{\end{eqnarray}}
\newcommand{\no}{\nonumber}    

\newcommand{\be}{\begin{eqnarray}}
\newcommand{\ee}{\end{eqnarray}}

\begin{document}
\title{\Large  \color{rossoCP3} ~~\\ LUX Constraints on Magnetic Dark Matter \\ in the $S\overline{E} \chi$y Model with(out) Naturality}
\author{Mads T. Frandsen}
\email{frandsen@cp3-origins.net} 
\author{Francesco  Sannino}
\email{sannino@cp3-origins.net} 
\author{Ian M. Shoemaker}
\email{shoemaker@cp3-origins.net} 
\author{Ole Svendsen}
\email{svendsen@cp3-origins.net} 
\affiliation{{\color{rossoCP3} CP$^{3}$-Origins} \& Danish Institute for Advanced Study {\color{rossoCP3} DIAS}, University of Southern Denmark, Campusvej 55, DK-5230 Odense M, Denmark}

\begin{abstract}
We study phenomenological constraints in a simple $S\overline{E} \chi$y extension of the Standard Model (SM) with a 125 GeV Higgs, a vectorlike heavy electron $(E)$, a complex scalar electron $(S)$ and a standard model singlet Dirac fermion $(\chi)$. 
The interactions among the dark matter candidate $\chi$ and the standard model particles occur via loop-induced processes involving the Yukawa interaction $S\overline{E} \chi$y. The model is an explicit perturbative realization of so-called magnetic dark matter.  The field content allows for a cancelation of quadratic divergences in the scalar masses at one-loop, a phenomenon which we refer to as perturbative naturality.  The basic model is constrained dominantly by direct detection experiments and its parameter space can be nearly entirely covered by up-coming ton-scale direct detection experiments. We conclude this work by discussing different variations of the model. 
\\[.1cm]
{\footnotesize  \it Preprint: CP3-Origins-2013-048 DNRF90, DIAS-2013-48}
 \end{abstract}

\maketitle
  
\section{Introducing the model}
After the discovery~\cite{Aad:2012tfa,Chatrchyan:2012ufa} of a new resonance with properties similar to the standard model (SM) Higgs particle, it is relevant to investigate minimal perturbative and non-perturbative extensions of the SM that can include dark matter candidates. 

The $S\overline{E}\chi \lowercase{y}$ model \cite{Dissauer:2012xa} is a perturbative example. In addition to a 125 GeV Higgs $H$, the model includes a vector-like heavy electron ($E$), a complex scalar electron ($S$) and a SM singlet Dirac fermion ($\chi$) as dark matter (DM) interacting via the Yukawa operator $S\overline{E}\chi \lowercase{y}$. 

In this paper we focus on the DM phenomenology and the {\it naturality} of the model, assuming the new states to be elementary and the DM a thermal relic \footnote{If the Higgs sector of the SM is (strongly) natural and described by new strong dynamics we can imagine this new sector as an effective description with $H, S, E, \chi$ being composite, e.g. \cite{Sannino:2008nv,Frandsen:2011kt}. The first lattice simulations of minimal models of composite DM appeared in \cite{Lewis:2011zb,Hietanen:2012qd,Hietanen:2012sz,Hietanen:2013fya,Hietanen:2013gva}.}.  The phenomenological analysis is performed without assuming any degree of naturality of the theory.   However, to gain insight on some more natural values of the couplings we will require perturbative naturality, in the form of the vanishing of the quadratic divergences of the scalars $H,S$ masses  \cite{Decker:1979cw,Veltman:1980mj}. {}We refer to \cite{Jack:1989tv,Antipin:2013exa} for a general classification and examples of the various possible extensions of the SM with respect to various degrees of naturality and associated predictive power. Furthermore, setting aside naturalness, the vacuum stability of the model and its interplay with gravitational corrections near the gravitational transition scale have been investigated in \cite{Antipin:2013bya}.  As we shall see the model is dominantly constrained by current direct detection experiments, in particular LUX \cite{Akerib:2013tjd} while constraints from LHC on the invisible width of the Higgs are subdominant in the considered parameter range. 

The model Langrangian features, in addition to the SM one, the following renormalizable terms \cite{Dissauer:2012xa}:
\begin{eqnarray}
\label{Sexy}
    \mathcal{L}_{S\overline{E}\chi {\rm y}}& =&  \mathcal{L}_{SM} + \bar{\chi}i\slashed{\partial}\chi - m_\chi\bar{\chi}\chi + \overline{E}i\slashed{D}E - m_E\overline{E}E - (S \overline{E}{\chi}  y + \text{h.c.} )
    \nonumber \\
& +& D_\mu S^{\dagger}D^\mu S - m_S^2S^{\dagger}S-\lambda_{HS}H^{\dagger}HS^{\dagger}S - \lambda_S (SS^{\dagger})^2\ ,\end{eqnarray}
where $D^\mu=\partial^\mu-ie\frac{s_w}{c_w}Z^{\mu}+ieA^\mu$, $s_w$ and $c_w$ represent the sine and cosine of the Weinberg angle. The gauge and self coupling for the Higgs come from: 
\begin{eqnarray}
D_\mu H^{\dagger}D^\mu H - \mu^2 H^{\dagger}H  - \lambda (H^{\dagger}H)^2 \ .
\end{eqnarray}

We assume the new couplings $y, \lambda_{HS}$ and $ \lambda_S$ are real and the bare mass squared of the $S$ field is positive so that the electroweak symmetry breaks via the SM Higgs doublet ($H$). The interactions among $\chi$, our potential dark matter candidate, and the SM fields occur via loop-induced processes involving the $S\bar{E} \chi$y operator in \eqref{Sexy}.

A similar model with $S,E$ both $SU(2)_W$ doublets and a splitting of the dark matter fermion into different Majorana mass eigenstates was considered in \cite{Weiner:2012gm} while related but more minimal models with EW singlet $S,\chi$ is discussed in \cite{Kim:2008pp,Fairbairn:2013uta,Masina:2013wja,Antipin:2013exa}.  We discuss simple variations of the model at the end of the paper.

\section{Relic density of Dark $S\overline{E}\chi \lowercase{y}$}
We being our study by assuming $\chi$ constitutes the DM of the universe and that the DM abundance is determined from thermal freeze-out of $\chi$ when its annihilation rate drops below the expansion rate of the Universe.

\subsection{Annihilation cross-sections in the early Universe}
 
There are two options for the annihilation channels: 1) DM can annihilate through one-loop induced interactions to SM final states; or 2) DM can annihilate directly at tree-level to $E^{+}E^{-}$ pairs through the $S\overline{E}\chi \lowercase{y}$ operator. In both cases of course, a given channel will only be open (or at least not strongly suppressed) if the DM mass exceeds the mass of the final state particles. We will assume that the heavy electron can mix with the SM-leptons and decay.

Annihilation into SM final states can occur via $s$-channel photon, $Z$ or Higgs exchange 
\be    
\mathcal{L}_{\chi-SM} = \frac{\lambda_{\chi}}{2}\overline{\chi}\sigma_{\mu \nu}\chi F^{\mu\nu} - \frac{s_{w}}{c_{w}} \frac{\lambda_{\chi}}{2}\overline{\chi}\sigma_{\mu \nu}\chi Z^{\mu\nu} + \lambda_{h\chi} h~ \overline{\chi}\chi
\ee
where  the magnetic dipole moment $\lambda_{\chi}$, and induced $\chi$-Higgs coupling are given respectively by 
\be \lambda_{\chi}(q^{2}) = \frac{F_{2}(q^{2})e}{2m_{\chi}},~~~~~ \lambda_{h\chi} = vy^{2} \lambda_{HS} \mathcal{A}_{h\chi}
\ee
where $F_{2}(q^{2})$ is the electromagnetic form factor, and the loop-factor $\mathcal{A}_{H\chi}$ is given in~\cite{Dissauer:2012xa}. When $m_{S} = m_{E} \gg m_{\chi}, m_{h}$ the two couplings are well-approximated by 
\be
\label{eq:effdmcoupl}
\lambda_{\chi}(0) \simeq \frac{e y^{2}}{32\pi^{2} m_{S}},~~~~~\lambda_{h\chi} \simeq \frac{vy^{2}\lambda_{HS}}{32 \pi^{2} m_{S}}.
\ee

\subsubsection{The magnetic photon/Z annihilation process}
The annihilation via the above Z and photon dipole interactions lead to \cite{Heo:2009vt}
\begin{align}
\sigma^{\gamma/Z}_{\chi \bar\chi \rightarrow f \bar f} \ v_\text{rel} &= \lambda^2_{\chi}  \frac{N_C^f \alpha \beta_f}{3} 
\big [(Q_f P_\gamma - \frac{v_f P_Z}{2 \cos^2 \theta_W} )^2 (s+2 m_f^2) + (\frac{a_f P_Z }{2 \cos^2 \theta_W}) (s-4 m_f^2)\big]
(s+8 m_\chi^2)
\\
\sigma^{\gamma/Z}_{\chi \bar\chi \rightarrow WW} \ v_\text{rel} &= \frac{\lambda^2_{\chi} \alpha \beta_W s}{24} (P_\gamma - P_Z)^2 \frac{s^2+20 s m_W^2 + 12 m_W^4}{m_W^4} (s+8 m_\chi^2)
\\
\sigma^{\gamma/Z}_{\chi \bar\chi \rightarrow ZH} \ v_\text{rel} &= \frac{\lambda^2_{\chi} |\vec{p}|}{2\sqrt{s}} (\frac{m_Z^2}{m_W \cos_\theta^2} P_Z)^2 (1+\frac{|\vec{p}|^2}{3 m_Z^2 })(s+8 m_\chi^2)
\end{align} 
where $N_C^f$ is a color factor with $N_C^\ell = 1$ for leptons and $N_C^q = 3$ for quarks,  $\beta_{i} = \sqrt{1- 4m_{i}^{2}/s}$, $Q_f$ is the charge of the fermion and $a_f, v_f$ are the ratios of vector and axial-vector couplings of the fermions to the $Z$ boson: $a_f=T^3_f, v_f = T^3_f-2Q_f \sin\theta_W^2$. 

Finally $P_\gamma={s^{-1}}$ and $P_Z=\left((s-m_Z^2)+i m_Z \Gamma_Z\right)^{-1}$ are the propagators of the photon and the $Z$-boson respectively and $\vec{p}$ is the momentum of the $Z$ (or $H$).

Note that in isolation both the photon and $Z$ contribution to the $\chi \chi \to WW$ annihilation channel grow with the energy, but their sum does not. Also, the quoted annihilation cross-sections  into $WW$ and $ZH$ are via the photon and $Z$ moment interaction respectively. The decay modes into $ZZ$ and $\gamma\gamma$ via two magnetic moment vertices are suppressed by two additional powers of $\lambda_\chi m_\chi$ (with the $m_\chi$ factor appearing due to helicity) and remain subdominant decay modes.

\subsubsection{The Higgs portal annihilation process }
The annihilation via the Higgs moment to SM final states gives (e.g \cite{Kim:2008pp}):
\begin{align}
 \sigma_{\chi \bar\chi}^h \ v_\text{rel} = \frac{\lambda_{h\chi}^{2}\beta_{\chi}^{2}}{16\pi \left[ (s-m_h^2)^2+(\Gamma_h m_h)^2 \right]} 
\left( \sum_{V=W,Z} s_V \left( \frac{2 m_V^2}{v}\right)^2 \left(2+\frac{(s-2 m_V^2)^2}{4 m_V^4}\right)\beta_V 
+ \sum_{f} N_C^f y_f^{2} \, s \, \beta_f^{3} 
\right)
\end{align}
where $s_{Z} = 1/2$ and $s_{W} = 1$, $\beta_{i} = \sqrt{1- 4m_{i}^{2}/s}$, and $y_{f}$ are the SM Yukawa couplings. 

\subsubsection{The t-channel annihilation process via $S$}

We finally consider the case in which $\chi$ is sufficiently heavy to annihilate into
 the heavy electron $E$ through a $t$-channel exchange of the $S$ scalar, which can occur when $m_{\chi} > m_{E}$.  The associated cross section is an $s$-wave and reads:
\bea \langle \sigma_{\chi \overline{\chi}}^{S} v_{rel} \rangle &=& \frac{y^{4} \left(m_{E}+ m_{\chi}\right)^{2}}{8 \pi \left(m_{E}^{2}-m_{X}^{2}-m_{S}^{2}\right)^{2}} \sqrt{1-\frac{m_{E}^{2}}{m_{\chi}^{2}}}+ \mathcal{O}(v^{2}) .
\eea
The lower limit on the mass of $E$, $m_{E} > 393$ GeV~\cite{Dissauer:2012xa}, implies that this annihilation channel will only be relevant for heavy DM.

 \subsection{Relic Density Calculation}
 To treat the thermal relic abundance properly near the Higgs resonance, we follow~\cite{Gondolo:1990dk} and carry out the thermal averaging of the cross section explicitly, yielding
 \be 
 \langle \sigma v_{rel} \rangle = \int_{4m_{\chi}^{2}}^{\infty} ds~ \frac{s \sqrt{s-4m_{\chi}^{2}}~K_{1}\left(\sqrt{s}/T \right)}{16 T m_{\chi}^{4} K_{2}^{2}\left(m_{\chi}/T\right)}~ \sigma v_{rel} , 
 \ee
where $K_{n}(x)$ is a modified Bessel function of the $n$'th kind. This is used to iteratively solve for the freeze-out temperature, $T_{F}$
\be x_{F}\equiv \frac{m_{\chi}}{T_{F}} = \log\left(\frac{m_{\chi}}{2\pi^{3}}\sqrt{\frac{45 M_{P}^{2}}{2g_{eff}x_{F}}} \langle \sigma v_{rel} \rangle \right)
\ee
and finally compute the relic abundance via
 \be \Omega_{DM}h^{2} = \frac{1.07 \times 10^{9}x_{F}}{\sqrt{g_{eff}}M_{p}\langle \sigma v_{rel} \rangle} , 
 \ee
 where $g_{eff}(T)$ are the effective energy degrees of freedom, and $M_{P}$ is the Planck mass. We use the values of $g_{eff}(T)$ given in the DarkSUSY code~\cite{Gondolo:2004sc} which include the improved QCD equation of state from~\cite{Hindmarsh:2005ix}.

\section{Phenomenology} 
 
 Before we examine the detailed phenomenology from recent experiments it is useful to asses the status of the model based on LEP data. First all DM models must be consistent with the observed invisible decay width of the $Z$ boson, which agrees well with the SM value. This coupling arises in our model at 1-loop and can be relevant when it is kinematicically accessible, $m_{\chi}< M_{Z}/2$.  The constraint from the invisible $Z$ width was investigated in~\cite{Dissauer:2012xa} where it was found that the constraint is always well-satisfied for $y < 4\pi$. 
 
 Furthermore, constraints arising from contributions to the oblique parameters are also easily satisfied by the scalar $S$ and the fermion $E$. Corrections to $S$, $T$, and $U$ in the presence of heavy scalars~\cite{Zhang:2006vt} and vector-like fermions~\cite{Zhang:2006de} vanish at 1-loop for $SU(2)$ singlets. 
 
Let us now examine the phenomenology of the model arising from recent experimental data.

\subsection{Direct Detection} 
 For the Higgs mediated interactions, we write the $\chi$-nucleon couplings as 
 \be f_{p,n} = \lambda_{h\chi} \frac{m_{p,n}}{v m_{h}^{2}} \left( \sum_{q=u,d,s}f_{Tq}^{p,n} + \frac{2}{9}f_{Tg}^{p,n}\right)
 \ee 
 where the coefficients $f_{Tq}^{p,n}$ , $f_{Tg}^{p,n} = 1- \sum f_{Tq}^{p,n}$ are taken from~\cite{Ellis:2000ds}.  To calculate the direct detection limits, we then compute the resulting bound following from the $\chi$-nucleus cross section
 \be \frac{d\sigma}{dE_{R}} = \frac{m_{N}}{2\pi v^{2}} \left( Zf_{p} + (A-Z) f_{n}\right)^{2} F^{2}(E_{R})
 \ee
 where $E_{R}$ is the nuclear recoil energy, $m_{N}$ is the nuclear mass, $v$ is the incoming DM velocity, and $F(E_{R})$ is the nuclear form factor. Note that in this case one can roughly estimate the constraint from simply rescaling the quoted limits on the spin-independent DM-nucleon cross section, $\sigma_{SI} = \left(\mu_{p}/\mu_{N}\right)^{2} \sigma_{N}/A^{2}$.

 Whereas in the case of the magnetic moment interaction, we follow~\cite{DelNobile:2012tx} and use the differential $\chi$-nucleus cross section
 \be 
 \frac{d \sigma}{dE_{R}} = \frac{\alpha_{EM}\lambda_{\chi}^{2}}{E_{R}} \left\{ \left[1-\frac{E_{R}}{v^{2}}\left(\frac{2m_{N}+m_{\chi}}{2m_{N}m_{\chi}}\right)\right] Z^{2}F^{2}(E_{R}) + \left(\frac{\bar{\lambda}_{nuc}}{\lambda_{p}}\right)^{2}  \frac{E_{R}}{v^{2}} \frac{m_{N}}{3m_{p}^{2}} F_{SD}^{2}(E_{R}) \right\},
 \ee
 where $\bar{\lambda}_{nuc}$ is an isotope weighted dipole moment of the nucleus and $F_{SD}$ is a form factor for the spin-dependent part of the scattering - the spin-dependent contribution to the scattering is negligible for the XENON experiments we are interested in here. For astrophysics, we will throughout assume a Maxwell-Boltzmann velocity distribution with a dispersion $v_{0} = 220$ km/s and an escape speed $v_{esc} = 544$ km/s. We will not explore the impact of astrophysical uncertainties in direct detection here, but note work in this direction~\cite{Green:2007rb,Green:2008rd,Peter:2009ak,Strigari:2009zb,McCabe:2010zh,Fox:2010bz,Fox:2010bu,Frandsen:2011gi,Kavanagh:2012nr,Friedland:2012fa,Frandsen:2013cna,DelNobile:2013cva,DelNobile:2013gba}.
 
The first data from LUX \cite{Akerib:2013tjd} was recently released with an exposure of $10,065$ kg--days. Our analysis follows that employed in~\cite{Cirigliano:2013zta}. The collaboration quotes an upper limit of 2.4 signal events for DM masses $< 10$ GeV \cite{lux-talk},  with up to 5.3 events allowed for larger masses. To be  conservative we apply a limit of 5.3 signal events to the whole mass range. To compute the number of signal events, we convolve the rate of nuclear recoils from DM scattering with the Poisson probability to produce the number of photoelectrons detected. We use the acceptance provided by \cite{Akerib:2013tjd}, 
and the energy-dependent absolute light-yield given in \cite{lux-talk}, with a sharp cutoff at 3 keV.
This procedure is found to reproduce the limits given by the collaboration.   
 
We also include projections for the future of direct detection that will be achieved in the near-term. First we mock-up a future LUX sensitivity by simply assuming all detector details remain the same but that the exposure is a factor 5 times larger than LUX's existing exposure. This will  be achieved with $\sim 1$ year of data. This is shown in each panel of Fig.~\ref{fig:phenorelic} as the dashed green line.  

Next, we estimate the sensitivity of XENON1T~\cite{Aprile:2012zx} with a 3 tonne-year exposure. Here we use a simplified treatment by taking the acceptance to be a flat 45$\%$ in the 2-30 keV energy window, and zero elsewhere. This is similar to other existing projections of XENON1T in the literature~\cite{Pato:2010zk}.  We exclude at 90$\%$ CL cross sections that yield more that 2.3 signal events, using Poisson statistics under the assumption that the collaboration reaches their goal of  $< 1$ background event in the fiducial region.

\subsection{Higgs Invsisible Decays}

The Higgs DM coupling given in Eq.~(\ref{eq:effdmcoupl}) also gives rise to an invisible decay width of the Higgs into the Dirac DM 
\begin{equation}
\Gamma = \frac{m_h}{8\pi}\lambda_{h\chi}^{2} \beta_\chi^3 
\end{equation}
For Majorana DM the decay width is half of the above result. From the results of \cite{Giardino:2013bma}, one deduces the $2\sigma$ estimate
\begin{equation}
{\rm Br}_{\rm inv} \simeq \frac{\Gamma[H\to 2\chi]}{\Gamma_{\text{SM}}+ \Gamma[H\to 2\chi] } \leq 0.25 \ , 
\end{equation}
where the first $\simeq$ is a very good approximation in our model since the only decay mode that is modified compared to the SM Higgs is the two-photon decay mode.

To get a rough upper(lower) bound on $\lambda_\chi(m_S)$ we take $\Gamma_{\text{SM}} \sim \Gamma_{b\bar{b}} \gg \Gamma_{\chi\bar{\chi}}$, $m_\chi \ll m_h$ and thus get 
\begin{equation}\frac{\lambda_{h\chi}^2}{3 \lambda_b^2 } < 0.25 \ , \qquad  {\rm i.e.} \qquad \frac{y^2  \lambda_{HS}}{\pi^2 } v< m_S \ .
\end{equation}
For $O(1)$ couplings this limit, applicable for $m_\chi < m_h$, allows $m_S$ well below the weak scale while for the coupling values we will be interested $y,\lambda_{SH}$ of the order of a few the limit is of the order of a TeV. In Fig.~\ref{fig:phenorelic} we show the limit without these approximations as the blue curves.

\subsection{Higgs decays to two-photons}
The Higgs decay width to two photons is modified by the presence of the new scalar state $S$ which interacts with both the Higgs and the photon. At one-loop level the decay width is
\begin{equation}
  \label{eq:hwidth}
    \Gamma[h\to \gamma\gamma]=\frac{\alpha_{EW}^2G_Fm_h^3}{128\sqrt{2}\pi^3}\left|\sum_iN_{c,i}Q^2_iF_{f,i}+F_W+\left(\frac{\lambda_{HS} v^2}{2 m_S^2}\right)F_S\right|^2,
\end{equation}
where $N_c$ is the number of colors and $Q$ is the charge of a given particle contributing to the process. Using the notation $\tau=\frac{4m^2}{m_h^2}$, the loop functions $F_i$ for fermions, bosons and scalars are  
\begin{equation}
  \label{eq:loopF}
  \begin{split}
    F_W&=2+3\tau+3\tau(2-\tau)f(\tau),\\
    F_f&=-2\tau(1+(1-\tau)f(\tau)),\\
    F_S&=-\tau(1-\tau f(\tau)),
  \end{split}
\end{equation}
where
\begin{equation}
  \label{eq:ftau}
  f(\tau)=
  \begin{cases}
    \left(\arcsin{\sqrt{1/\tau}}\right)^2 & \text{if}\quad \tau \geq 1 \\
    -\frac{1}{4}\left[\log\left(\frac{1+\sqrt{1-\tau}}{1-\sqrt{1-\tau}}\right)-i\pi\right]^2 & \text{if} \quad\tau < 1.
  \end{cases}
\end{equation}

\begin{figure*}[t!]
  \centering
  \includegraphics[width=0.48\textwidth]{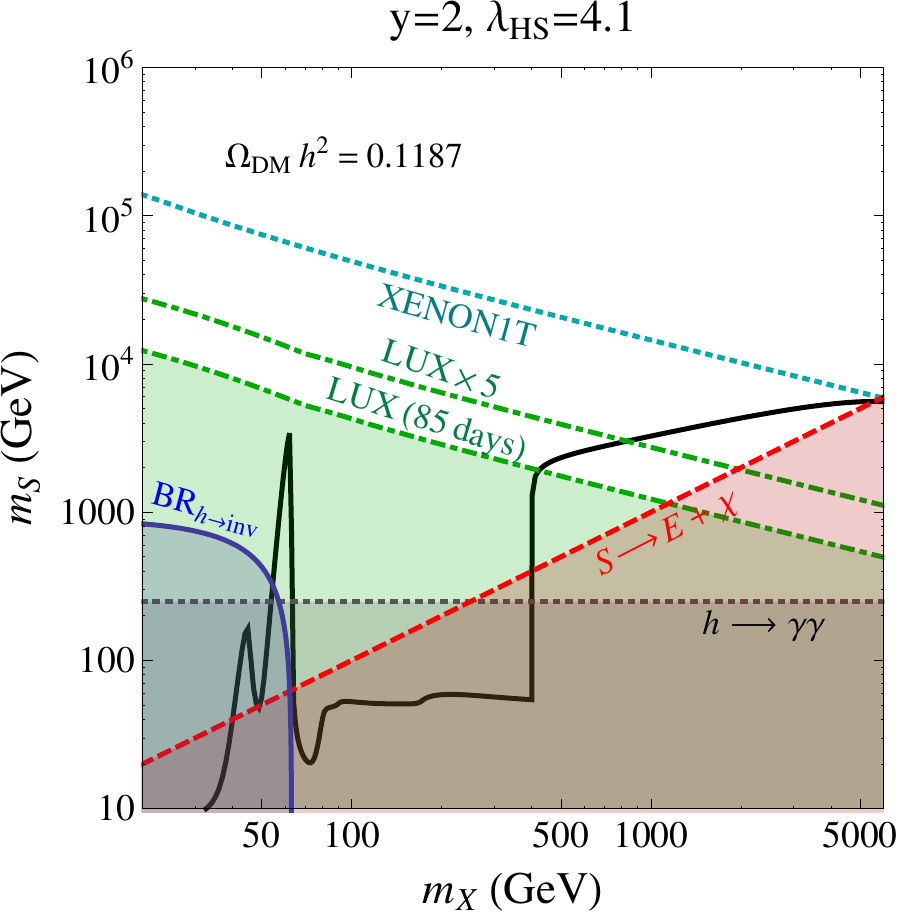}
    \includegraphics[width=0.48\textwidth]{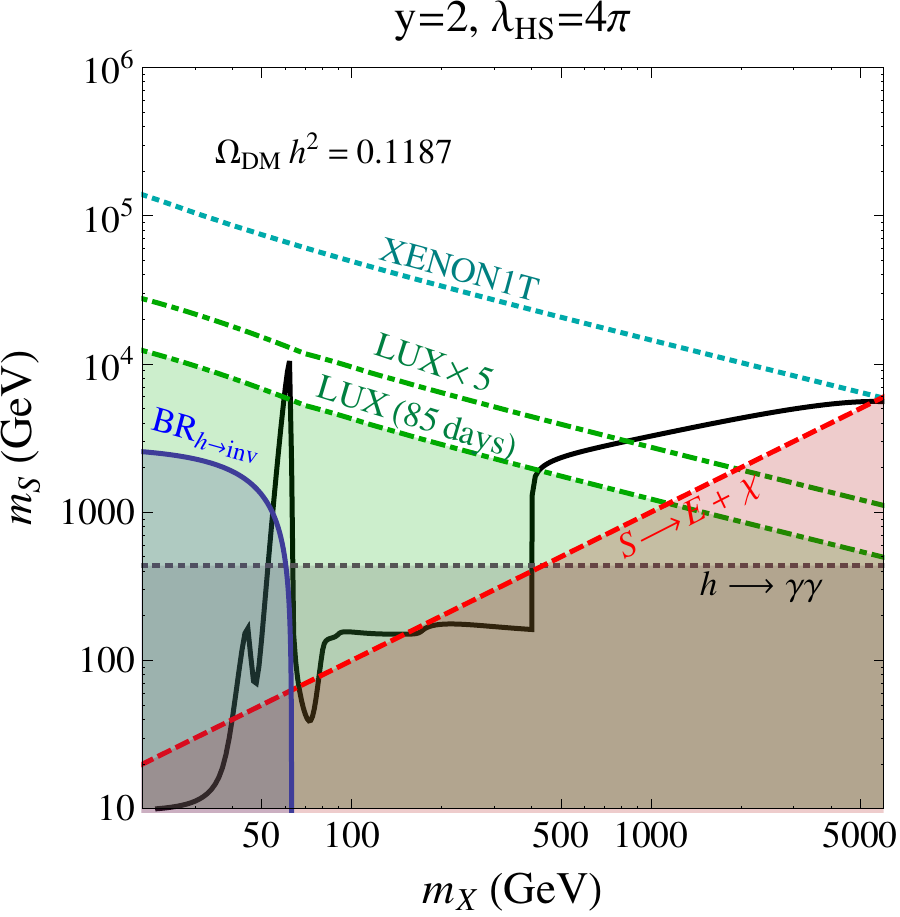}
  \caption{Detailed comparison of constraints on the $S\overline{E} \chi$y model for $y=2$, $m_{E} =400$ GeV, for $\lambda_{HS}$ = 4.1 (left panel) and $4\pi$ (right panel). We display in red the constraint arising from the requirement that $S$ decays, $S\rightarrow E + \chi$ where $E$ decays further. Blue shows the limit from requiring BR$\left(h\rightarrow {\rm inv}\right) \le 25$\%, whilst green shows the LUX limit arising from the Higgs portal interaction. Black dotted shows the limit from the ATLAS/CMS limit on $h\rightarrow \gamma \gamma$. We also indicate the sensitivity of LUX with a factor 5 increase in exposure, and a XENON1T projection. See text for analysis details.
  }
  \label{fig:phenorelic}
\end{figure*}

Approximating the relevant loop-factors by $F_W+F_t\simeq -6.5$, $F_S\simeq 0.34$ their asymptotic values, the ratio of the di-photon partial width to the SM-Higgs one becomes
\begin{equation}
\mu_{\gamma \gamma} = \frac{\Gamma[h\to \gamma\gamma]}{\Gamma[h^{\rm SM}\to \gamma\gamma]} \sim (1-\frac{\lambda_{HS} v^2}{20 m_S^2})
\end{equation}
while the currently measured values are: $\mu_{\gamma \gamma}^{\rm ATLAS}=1.33^{+0.21}_{-0.18}$ \cite{Aad:2013wqa} and $\mu_{\gamma \gamma}^{\rm CMS}=1.05\pm 0.36$ \cite{CMS-PAS-HIG-13-016}.
Thus from a naive combination we 
require $m_S > \frac{\sqrt{\lambda_{HS}}}{2} v$ within 2$\sigma$.
Unless the quartic coupling is near the perturbative limit $\lambda_{HS} \sim 4\pi$ the reduction of the Higgs width is small compared to the current limit. Moreover from Fig.~\ref{fig:phenorelic} it follows that the modification is negligible compared to the expected sensitivity of LHC in the parameter regions where the $S\overline{E} \chi$y model can provide the correct relict density and is not yet ruled out by data.

\subsection{Results}
We present the combined constraints in Figure~\ref{fig:phenorelic}. 
The remaining viable regions are near the Higgs pole and in a 2-5 TeV window dominated by annihilation into $E^{+}E^{-}$ pairs.   

The main tension arises from the very strong limits from direct detection. While the Higgs portal interaction for annihilation depends on the coupling $\lambda_{HS}$, the direct detection sensitivity is dominantly determined by the photon dipole interaction, and is thus insensitive to $\lambda_{HS}$. Thus the LUX/XENON1T and relic abundance curves can be shifted relative to each other from varying $\lambda_{HS}$. The right panel of Fig.~\ref{fig:phenorelic} demonstrates that unless $\lambda_{HS}$ is $\sim 4\pi$, the Higgs resonance region will not be a viable option for the thermal relic density.

\section{Improving on naturalness}
So far we have concentrated on the phenomenological aspects of the $S\overline{E} \chi$y extension of the SM with emphasis on the DM phenomenology. We learned (see Fig.~\ref{fig:phenorelic}) that large values of the portal coupling, while still remaining within the perturbative regime, open the window of the DM parameter space compatible with experiments.  We will momentarily see that these values of the portal coupling are, in fact, predicted by imposing the one-loop vanishing of the quadratic divergences of the scalar-masses in the theory. 

From the Lagrangian in \eqref{Sexy} and following \cite{Antipin:2013exa} we determine the following quadratically divergent terms for the physical Higgs state $h$ and extra scalar $S$ two point functions:
\begin{eqnarray}
\delta m_h^2&=&\left(-6 y_t^2 +\frac{3}{4}(g^2+{g'}^2) + \frac{3}{2} g^2 +6 \lambda + \lambda_{HS}\right) \frac{\Lambda^2}{16\pi^2} \ , \no
\\
\delta m_S^2&=&\left(-2 y^2 +2 \lambda_{HS}  + 3{g'}^2 + 4 \lambda_S\right) \frac{\Lambda^2}{8\pi^2} \ .
\label{VCs}
\end{eqnarray}
Here $\Lambda$ is some given energy scale, higher than the electroweak scale, above which the present description is modified. The couplings are, however, evaluated at the electroweak scale. Because the top-Yukawa and the Higgs self couplings are experimentally known requiring these two equations to vanish \cite{Decker:1979cw,Veltman:1980mj} leads to the constraints: 
\begin{equation}
\lambda_{HS} \sim 4 \ , \qquad  y\sim \sqrt{4+2\lambda_S} \ .
\label{constraints} 
\end{equation}
 Furthermore stability of the theory requires $\lambda_S \geq 0$ implying from \eqref{constraints} $y \gtrsim 2$. 
The perturbative expansion is in $\alpha_{y_t} = y_t^2/4\pi$, $\alpha_{y} = y^2/4\pi$, $\alpha_{\lambda} = \lambda/4\pi$,  $\alpha_{\lambda_{HS}} = \lambda_{HS}/4\pi$ and $\alpha_{\lambda_S} = \lambda_S/4\pi$ insuring that we remain within the perturbative regime of the theory. 

Requiring cancellation of the quadratic divergences of the scalar operators renders, at least in perturbation theory, the electroweak scale more stable, and the full theory (according to the renormalization group analysis) more natural. This is because scalar mass terms are relevant operators driving the theory towards the highest scale in the problem, typically the scale of new physics or, in absence of the latter, the gravity transition scale \cite{Antipin:2013bya}. A general classification of the various shades of naturality appeared recently in \cite{Antipin:2013exa}. Because naturality is ensured perturbatively we are not allowed to investigate energy scales much higher than the ones accessible within a perturbative renormalization group approach. 

We have associated the concept of degrees of naturality of a theory to the absence or to the softening, at the quantum level, of relevant operators destabilizing the low energy physics scale. The virtue of this mathematical definition is that it is free from ambiguities being deeply rooted in the renormalization group concept.

\begin{figure}[t]
  \centering
    \includegraphics[width=0.42\textwidth]{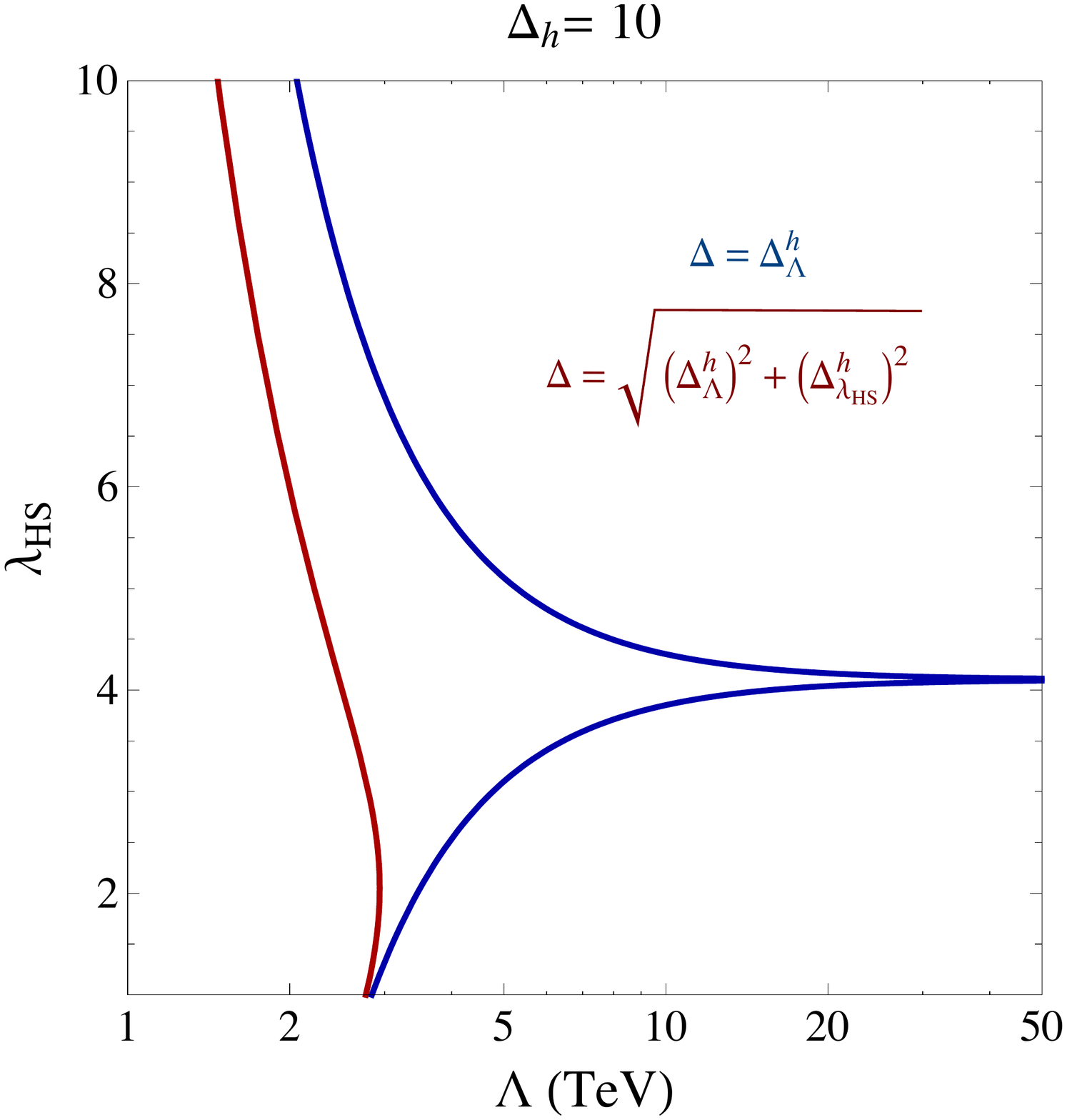}
        \includegraphics[width=0.42\textwidth]{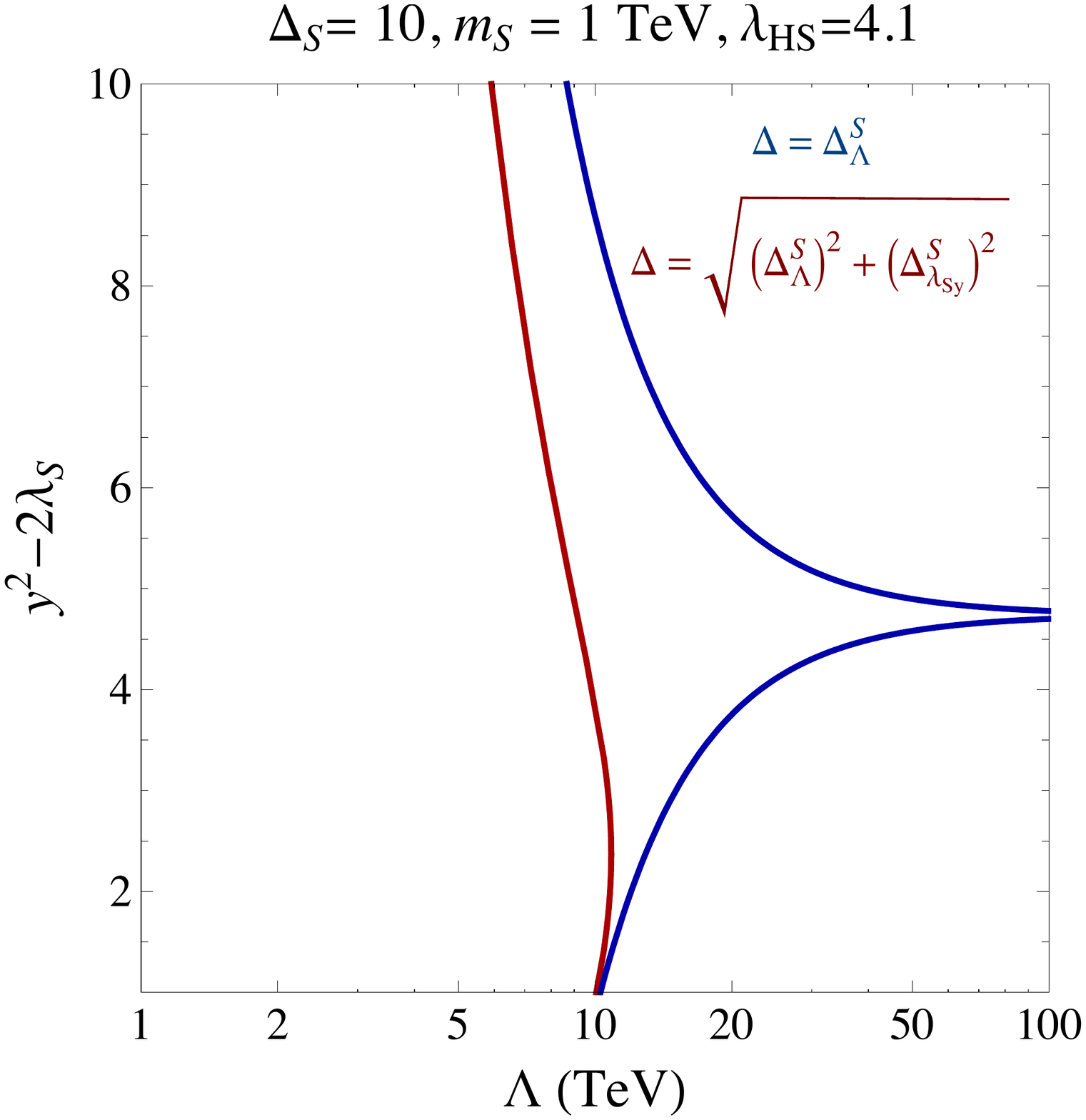}
  \caption{An illustration of the impact of different fine-tuning definitions for $\Delta_{h} = 10$ (left panel) and $\Delta_{S} = 10$ (right panel).
  }
  \label{fig:ftI}
\end{figure}

In addition to the concept of naturality, however, one can also discuss the more delicate issue of fine-tuning of a theory with respect to the stability of the electroweak scale. Following  \cite{Ellis:1986yg,Barbieri:1987fn}, for example, for each unknown parameter $p_i$ of the theory one can determine $\Delta_i$  as follows
\begin{equation}
\label{Eq:FTmeasure}
\Delta_i \equiv \Big|  \frac{p_i}{m_Z^2} \frac{\partial m_Z^2}{\partial p_i} \Big| = \Big| \frac{\delta m_Z^2}{m_Z^2} \Big| \ .
\end{equation}
Here $m_Z$ is the $Z$ mass and represents the electroweak scale. The degree of fine-tuning is then defined to be the quantity $\Delta$ given by either one of the following definitions
\begin{equation}
 \Delta \equiv {\rm Max} \left\{\Delta_i \right\} \ , \qquad {\rm or} \qquad  \Delta \equiv \sqrt{\sum_i \Delta_i^2} \ .
\end{equation}
 For example $\Delta=10$ would imply a tuning of one part in ten of $m_Z$. One can, in principle, use as reference scales for our theory $m_h^2$, and $m_S^2$ rather than $m_Z^2$.  We assume, for simplicity, $\lambda, y_t$ to have a fixed value and vary $\Lambda$ (the scale of new physics), $\lambda_{HS}$ and $\lambda_{Sy}\equiv y^2-2 \lambda_S$ and determine $\Delta^h$ and $\Delta^S$ given by:
\begin{equation}
\Delta^h \equiv \sqrt{(\Delta_\Lambda^{h})^2 + (\Delta_{\lambda_{HS}}^h)^2} \ , \qquad 
\Delta^S \equiv \sqrt{(\Delta_\Lambda^{S})^2 + (\Delta_{\lambda_{Sy}}^S)^2} \ .
\label{Fine}
\end{equation}
The conditions used in \eqref{VCs}, by construction, naturally minimize the dependence on the cutoff scale $\Lambda$ and therefore the quantities  $\Delta_\Lambda^{h}$ and $\Delta_\Lambda^{S}$. In particular, for the values of the couplings given in \eqref{constraints} these two terms vanish. In other words, to the one-loop level, we are not sensitive to the variation of the cutoff scale $\Lambda$ assuming the relations given in \eqref{constraints} are valid at the electroweak scale. We illustrate in Fig.~\ref{fig:ftI} the relation between the value of the couplings at the electroweak scale and the cutoff assuming a tuning of  $\Delta_\Lambda^{h} =\Delta_\Lambda^{S} = 10$. Clearly the apparent independence of the theory on $\Lambda$ when assuming \eqref{constraints} is not valid to higher orders and therefore the associated fine-tuning analysis must be taken with a grain of salt.

We now turn to the definition of fine-tuning given in \eqref{Fine}. If we require $\Delta_h = \Delta_S =10 $ we obtain the (left-most) red curves of the two panels of Fig.~\ref{fig:ftI}. It is clear that the major contributions to the fine-tuning come, for large values of the couplings, from $\Delta_{ \lambda_{HS}}^h$ for the left panel and  $\Delta_{\lambda_{Sy}}^S$ for the right one. This is obvious from the definition of these quantities. Furthermore it is $\Delta_h$, associated to the lowest energy scale, to provide the strongest constraint on $\Lambda$ and therefore we can concentrate on this quantity. Our conclusions are therefore that if we assume perturbative (delayed) naturality, a well defined concept, the couplings are expected to be near the ones given in \eqref{constraints}. However if we ignore the renormalization group naturality argument and use instead the fine-tuning definition given in \eqref{Fine}, the highest cutoff scale we can access is $\Lambda \simeq 3$~TeV for $\lambda_{HS}$ around half of the value assumed requiring perturbative naturality. 
Higher order corrections will be considered elsewhere 


\section{Conclusions, variations on the theme, and a non-magnetic example}
We have shown that the mass of the DM candidate in the $S\overline{E} \chi$y model is constrained by direct detection experiments to be in the range $0.5-5$~TeV, as is clear from Fig.~\ref{fig:phenorelic}. 
The region of DM masses around half the Higgs mass is severely constrained, although not completely excluded yet for very large values of the Higgs portal coupling.  For the associated masses and couplings the model is beyond the reach of LHC sensitivity.

In the near future we expect stronger constraints to come from direct detection experiments that will surpass the present sensitivity of XENON100 and LUX. 
To rule out the Higgs resonance region, as well as the high mass region for large portal couplings, roughly a factor $\sim2$ improvement in the cross-section limit ($\sqrt{2}$ in $m_S$ limit) is needed --- still without taking into account uncertainties in astrophysical halo parameters. This should be nearly possible with 300 live days of LUX as expected at the end of 2015.

Throughout this study we assumed charge (equally hypercharge) assignments $Q(S)=Q(E)=-n e$ with $n=1$, though one could consider more general assignments $n \neq 1$. If $n$ is fractional, $E$ cannot decay and stable fractionally charged particles are very strongly constrained. If $n$ is an integer larger than one, the dominant direct detection cross-section is increased by $n^2$ while the dominant annihilation cross-sections, i.e. the Higgs pole and $t$-channel exchange of $S$ are unaffected. From Fig.~\ref{fig:phenorelic} it follows that $n \gtrsim 2$ are ruled out with perturbative couplings ($y^2<4\pi$, $\lambda_{HS}<4\pi$). We are thus left with $|n|=1$ for magnetically interacting dark matter. 

A different modification of the model that allows for a significant weakening of direct detection altogether is to introduce a mass splitting $\delta(\chi_L^2 + {\chi^*_R}^{2}) $ \cite{TuckerSmith:2001hy} between the two Weyl components of $\chi=(\chi_L, \chi_R)$, such that the magnetic moment becomes a transition magnetic moment between the new mass eigenstates.
Direct detection limits are significantly weakened even with a modest value of $\delta \sim O(100)$ keV \cite{TuckerSmith:2001hy}.

We conclude by discussing the case in which the scalar $S$ is a real singlet and $\chi$ is taken to be a SM singlet Weyl fermion. An intriguing version of this  model was recently analyzed in \cite{Antipin:2013exa} where  the model was required to simultaneously break the electroweak symmetry via the Coleman-Weinberg mechanism and the quadratic divergences for the scalars were set to zero, at the electroweak scale, as done for the $SE\chi$y case above. 

\begin{figure}[t]
  \centering
    \includegraphics[width=0.5\textwidth]{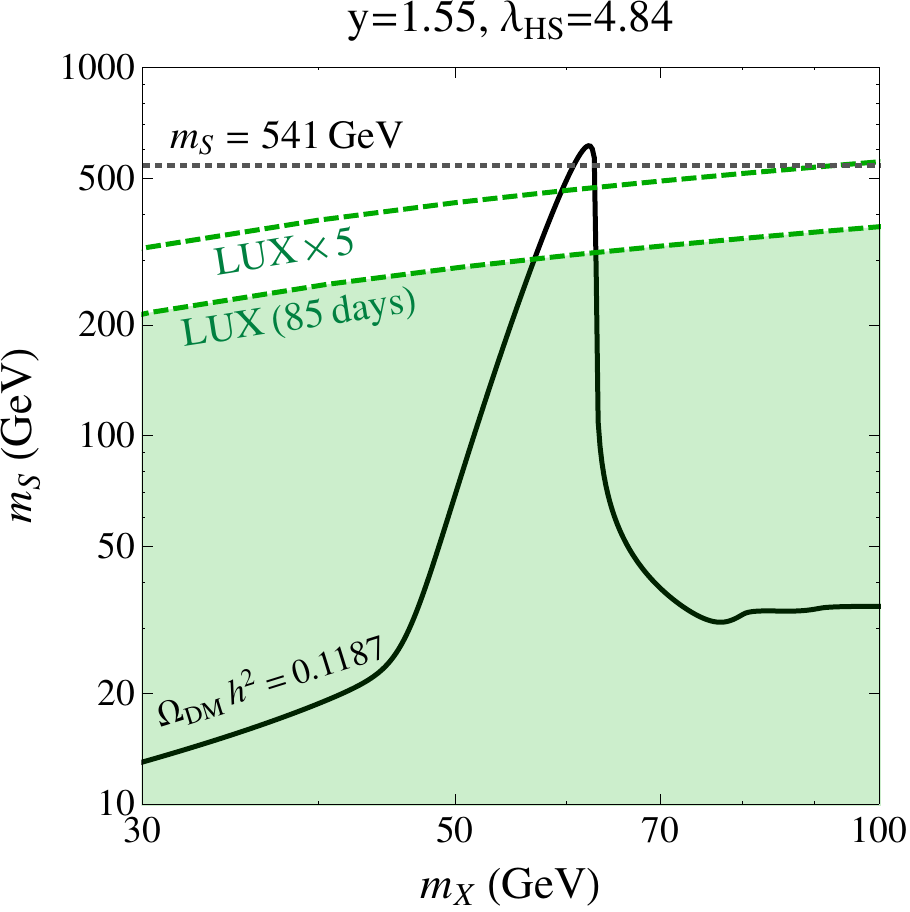}
  \caption{Here we plot the relic abundance (solid black) and direct detection constraints on the $S 
 \chi \chi$ model from~\cite{Antipin:2013exa}. Current LUX constraints exclude the shaded green region, while a LUX projected sensitivity with a factor 5 increase in exposure is indicated by the dashed green curve. The tree-level $S$ scalar mass $m_{S} = 541$ GeV arises as a prediction from the Veltman conditions as discussed in~\cite{Antipin:2013exa}. 
  }
  \label{fig:SXX}
\end{figure}
The model Lagrangian is given in Eq. (52) of \cite{Antipin:2013exa} where it was shown that the model predicts a one-loop generated mass for the Higgs of $126$~GeV and a tree-level mass for $S$ of about $540$~GeV.  Here we simply modify Eq. (52) of \cite{Antipin:2013exa} by adding an explicit Majorana mass term for the fermion of the model. It is straightforward to show that this addition does not change the predictions for the Higgs and $S$ masses, but allows to investigate $\chi$ as possible cold thermal-relic DM candidate. In this model the Coleman-Weinberg analysis requires, at the tree-level and at the electroweak scale, a vanishing of the Higgs self-coupling while the cancellation of the quadratic divergences fix the portal coupling to have a value of around $\lambda_{HS}\approx 4.84$. Simultaneously stability of the $S$ potential requires a Yukawa coupling of the $S\chi\chi$ interactions to be larger than about $y\approx 1.55$. 

Assuming these values of the couplings we demonstrate in Fig.~\ref{fig:SXX} that the model provides viable thermal relic dark matter, for a mass of $\chi$ near half the Higgs mass. Importantly, the direct detection limits are weaker than for the $S\overline{E} \chi$y model because here scattering on nuclei proceeds entirely through the Higgs portal, rather than the $S\overline{E} \chi$y magnetic moment. This also means that, in contrast with $S\overline{E} \chi$y, it is not possible to shift the LUX and relic abundance curves relative to each other since they both go through the same Higgs interaction. 

Intriguingly, we see from Fig.~\ref{fig:SXX} that the prediction for the S scalar mass of 541 GeV from \cite{Antipin:2013exa} is consistent both with the thermal relic and present direct detection constraints. Near-term improvements from direct detection will be able to fully test the remaining parameter space.

 \acknowledgments
We thank Levent Solmaz for collaboration on the initial stages of this work.  The CP$^{3}$
-Origins centre is partially funded by the Danish National Research Foundation,
grant number DNRF90.
MTF
acknowledges a `Sapere Aude' Grant no. 11-120829 from the
Danish Council for Independent Research.
 \bibliographystyle{ArXiv}
\bibliography{sexy.bib}

\end{document}